\documentstyle[PASJadd,psfig]{PASJ95}

\begin{document}

\title{Viscous Transonic Decretion in Disks of Be Stars}

\author{Atsuo T.\ {\sc Okazaki}
\\
{\it Faculty of Engineering,
                Hokkai-Gakuen University,
                Toyohira-ku, Sapporo 062-8605}\\
{\it E-mail: okazaki@elsa.hokkai-s-u.ac.jp}}

\abst{We study the characteristics of the outflow in disks of Be stars,
 based on the viscous decretion disk scenario. In this scenario, the
 matter ejected from the equatorial surface of the star drifts outward
 because of the effect of viscosity, and forms the disk. For
 simplicity, we adopt the $\alpha$-prescription for the viscous
 stress, and assume the disk to be isothermal. Solving the resulting
 wind equations, we find that a transonic solution exists for any
 value of $\alpha$. The sonic point is located at $r > 100 R$ for
 plausible values of parameters, where $R$ is the stellar radius. The
 sonic radius is smaller for higher temperature and/or larger
 radiative force. We also find that the topology of the sonic point is
 nodal for $\alpha \gtsim 0.95$, while it is of saddle type for
 $\alpha \ltsim 0.9$. We expect that the sonic point in the former
 case is unstable, whereas that in the latter case is stable. The
 outflow is highly subsonic in the inner part of the disk. Roughly,
 the outflow velocity increases linearly with $r$ and the surface
 density decreases as $r^{-2}$. Interestingly, the disk is near
 Keplerian in the inner subsonic region, while it is angular momentum
 conserving in the outer subsonic region and in the supersonic
 region. Our results, together with the observed range of the base
 density for Be star disks, suggest that the mass loss rate in the
 equatorial region is at most comparable with that in the polar
 region.}

\kword{hydrodynamics
            --- radial velocity
            --- stars: Be
            --- stars: mass loss
            --- stars: winds}

\maketitle
\thispagestyle{headings}

\section{Introduction}

 Be stars are non-supergiant early-type stars with Balmer emission
 lines, whose spectral types range from late O- to early
 A-type. Extensive studies have revealed that a Be star has a
 two-component extended atmosphere, a polar region and a cool ($\sim
 10^4\,{\rm K}$) equatorial disk. The polar region consists of a
 low-density, fast outflow emitting UV radiation. The terminal
 velocity of the polar wind is of the order of $10^3\,{\rm km}\,{\rm
 s}^{-1}$. The wind structure is well explained by so-called
 line-driven wind model, in which the radiative acceleration results
 from the scattering of the stellar radiation in an ensemble of
 spectral lines (Castor et al.\ 1975; Abott 1982).  The mass loss rate
 of this region inferred from the UV lines is about
 $10^{-11}-10^{-9}\,M_\solar\,{\rm yr}^{-1}$ (Snow 1981).

 On the other hand, the equatorial disk consists of a high-density
 plasma from which the optical emission lines and the IR excess
 arise. The radial velocity of the disk is smaller than a few ${\rm
 km\,s}^{-1}$, at least within $\sim 10$ stellar radii (Hanuschik
 1994, 2000; Waters, Marlborough 1994). The small radial velocity and
 the small pressure gradient force, as in viscous accretion disks,
 suggest that the equatorial disk is pressure-supported and
 geometrically thin in the vertical direction and is
 rotationally-supported in the radial direction. In fact, the presence
 of near Keplerian disks around Be stars is supported observationally
 as well as theoretically. For example, the relation between the size
 of the H$\alpha$-emitting region of Be disks resolved with the
 optical interferometer (Quirrenbach et al.\ 1997) and the separation
 of the double peaks of the H$\alpha$ line is in agreement with that
 expected for near Keplerian disks. In addition, the long-term
 line-profile variations are well explained by global disk
 oscillations with a one-armed density perturbation pattern, which can
 be present only in near Keplerian disks (Okazaki 1991; Papaloizou et
 al.\ 1992; Hanuschik et al.\ 1995). The mass loss rate through the
 disk is not known. The density in the disk derived from the IR excess
 and the observed upper limit of the radial velocity impose the upper
 limit of the mass loss rate of several $\times
 10^{-9}\,M_\solar\,{\rm yr}^{-1}$ (Waters, Marlborough 1994).

 At present, the viscous decretion disk model proposed by
 Lee et al.\ (1991) is the only model that naturally yields near
 Keplerian disks around Be stars. In this model, matter supplied from
 the equatorial surface of the star gradually drifts outward by the
 viscous torque and forms the disk. Assuming Keplerian rotation and
 neglecting the advective term in the equation of motion, Lee et al.\
 (1991) obtained steady and thermally-stable structure of viscous
 decretion disks around Be stars.

 Later, using a 3D Smoothed Particle Hydrodynamics code,
 Kroll and Hanuschik (1997) studied the evolution of the gas
 ejected temporarily from a Be star to model the transient disk
 formation/decay process around $\mu$~Cen. They found that the gaseous
 particles gradually expand and form a near Keplerian disk in the
 viscous timescale.

 These works show that the viscous torque is likely the agent that makes
 a disk near Keplerian. However, since the specific angular momentum
 in the near Keplerian disk increases with radius, Be disks are unlikely
 near Keplerian far from the star. A general belief is that the disk
 which is Keplerian near the star becomes angular momentum conserving
 far from the star, although no theoretical or observational study has
 confirmed this transitional feature of Be disks.

 The purpose of the present paper is to investigate whether Be disks
 have this transitional feature, based on the viscous decretion disk
 scenario. In section~2, we describe the disk model and formulate the
 eigenvalue problem for obtaining the radial structure of viscous
 decretion disks. We present the transonic solutions and discuss the
 topology of the sonic points in section~3. In section~4, we draw our
 conclusions and discuss several issues which remain unsolved.

\section{Basic Equations for Viscous Decretion Disks}\label{sec:model}

 We assume that the circumstellar disk of a Be star is steady,
 geometrically thin, and symmetric about the rotational axis
 and the equatorial plane.
 For simplicity, we adopt the Shakura-Sunyaev's prescription for the
 viscous stress.
 Since the deviation from a point mass potential due to
 the rotational distortion of the star is small in general,
 we neglect the quadrupole contribution to the gravitational potential.
  We use a cylindrical coordinate system
 $(r,\phi,z)$ to describe the disk.

 Millar and Marlborough (1998, 1999) computed the temperature
 distribution within $100R$ around the B0 star $\gamma$~Cas and the
 B8-9 star 1~Del by balancing at each position the rates of energy
 gain and energy loss, where $R$ is the stellar radius. They found
 that the disk temperature is roughly constant in the radial direction
 within $100R$. Since solving the hydrodynamical equations together
 with the energy equation is a formidable task, we assume for
 simplicity that the disk is isothermal. We believe that this
 assumption is a reasonable one as a first step to model the disk.

 Under these assumptions, the vertically-integrated
 equations describing the mass, momentum, and angular momentum
 conservation in the disk around the star of mass $M$, together with
 the equation of state and the Shakura-Sunyaev viscosity prescription,
 are
 \begin{equation}
    \dot{M} + 2 \pi r V_r \Sigma = 0,
       \label{eqn:mass}
 \end{equation}
 \begin{equation}
    -V_r {{{\rm d} V_r} \over{{\rm d} r}} + {V_\phi^2 \over r}
       - {{GM} \over r^2}
       - {1 \over \Sigma} {{{\rm d} W} \over{{\rm d} r}}
       + F_{\rm rad}
       + {{3 W} \over {2 r \Sigma}}
       = 0,
       \label{eqn:mom}
 \end{equation}
 \begin{equation}
    V_r {{{\rm d} V_\phi} \over{{\rm d} r}}
       + {{V_r V_\phi} \over r}
       - {1 \over{r^2 \Sigma}}
         {{\rm d} \over{{\rm d} r}} \left( r^2 t_{r \phi} \right) = 0,
       \label{eqn:angmom}
 \end{equation}
 \begin{equation}
    W = c_{\rm s}^2 \Sigma,
       \label{eqn:eos}
 \end{equation}
 \begin{equation}
    t_{r\phi} = -\alpha W,
       \label{eqn:alpha}
 \end{equation}
 where $\dot{M}$ is the decretion rate,
 $\Sigma$, $W$, and $t_{r \phi}$ are the vertically-integrated density,
 pressure, and $r$-$\phi$ component of the viscous stress, respectively,
 $V_r$ and $V_\phi$ are the radial and azimuthal components
 of the vertically-averaged velocity, respectively,
 $F_{\rm rad}$ is the vertically-averaged radiative force per unit mass,
 $c_{\rm s}$ is the isothermal sound speed,
 and $\alpha$ is the viscosity parameter.
  The last term of the left-hand-side of equation~(\ref{eqn:mom})
 is the correction for the decrease
 of radial component of the effective gravity away from the equator
 (e.g., Matsumoto et al.\ 1984).

 Since the radial flow in the inner part of viscous decretion disks
 is considered to be highly subsonic (Lee et al.\ 1991),
 the radiative acceleration would arise not from
 the optically-thick strong lines
 but from optically-thin weak lines
 (and from the optically-thin continuum).
 Since almost nothing is known about the form of the
 radiative force due to an ensemble of optically-thin lines,
 we adopt in this paper the parametric form
 used by Chen and Marlborough (1994):
 \begin{equation}
    F_{\rm rad} = \frac{GM\Gamma}{r^2}
            + \frac{GM(1-\Gamma)}{r^2}
              \eta \left( {r \over R} \right)^\epsilon,
    \label{eqn:Frad_gen}
 \end{equation}
 where $\eta$ and $\epsilon$ are parameters which characterize
 the force due to the ensemble of optically-thin lines.
 In the above expression,
 $\Gamma$
 is the Eddington factor
 that accounts for the reduction in the effective gravity
 due to electron scattering.
 In the rest of this paper,
 we neglect the Eddington factor $\Gamma$,
 because for typical values for Be stars,
 $\Gamma$ is as small as
 $\sim 0.03$ for a B0V star and $\sim 0.003$ for a B5V star.
 The radiative force is then written as
 \begin{equation}
    F_{\rm rad} \simeq \frac{GM}{r^2}
              \eta \left( {r \over R} \right)^\epsilon.
    \label{eqn:Frad}
 \end{equation}

 Eliminating $W$, $\Sigma$, and $t_{r\phi}$ from
 equations~(\ref{eqn:mass})--(\ref{eqn:angmom}),
 we obtain the following equations which describe
 the steady viscous flow:
 \begin{equation}
    \left( V_r - {c_{\rm s}^2 \over V_r} \right)
       {{{\rm d}V_r} \over {{\rm d}r}}
       = g_{\rm eff}
       + {\ell^2 \over r^3} + {5 \over 2}{c_{\rm s}^2 \over r},
    \label{eqn:windeq1}
 \end{equation}
 \begin{equation}
    \ell = \ell_{\rm s} + \alpha c_{\rm s}^2
              \left({r_{\rm s} \over c_{\rm s}}-{r \over V_r}\right),
    \label{eqn:windeq2}
 \end{equation}
 where $g_{\rm eff}$ is the effective gravity defined by
 \begin{equation}
    g_{\rm eff} = -{{GM}\over{r^2}} + F_{\rm rad}
         \simeq -{{GM}\over{r^2}}
         \left[1-\eta\left({r \over R}\right)^\epsilon\right],
    \label{eqn:geff}
 \end{equation}
 $\ell=rV_\phi$ is the specific angular momentum, and
 $r_{\rm s}$ and $\ell_{\rm s}$ are the radius of the sonic point
 and the specific angular momentum at the sonic point,
 respectively.
 It is important to note that
 the angular momentum distribution of the flow is not
 given a priori but obtained, with the radial velocity
 distribution, as the solution of equations~(\ref{eqn:windeq1}) and
 (\ref{eqn:windeq2}). In these equations,
 $r_{\rm s}$ and $\ell_{\rm s}$ are the eigenvalues.
  Note also that, from equation~(\ref{eqn:windeq2}), which indicates
 $\ell+\alpha c_{\rm s}^2r/V_r$ is constant, $V_r$ has to approximately
 linearly increase with $r$ in order for the disk to be near Keplerian.

 We solve equations~(\ref{eqn:windeq1}) and (\ref{eqn:windeq2}) by
 the shooting method with the starting point at $r=R$. Each trial
 integration needs two starting values, one of which is freely
 specifiable. We specify $V_r$ as a freely specifiable starting value
 and choose $\ell$ so as to be consistent with these equations in the
 very vicinity of $r=R$.

\section{Viscous Transonic Solutions}\label{sec:results}

 Wind equations~(\ref{eqn:windeq1}) and (\ref{eqn:windeq2}) include
 five independent parameters which characterize the solution.
 They are the viscosity parameter $\alpha$, the isothermal sound speed
 $c_{\rm s}$, the star's break-up velocity $(GM/R)^{1/2}$, and the
 radiative parameters $\eta$ and $\epsilon$. Since, in our model, the
 disk is assumed to be isothermal, the structure of the viscous
 transonic outflow is independent of the mass decretion rate $\dot{M}$
 [see section~3 of Abramowicz and Kato (1989)]. Solving
 equations~(\ref{eqn:windeq1}) and (\ref{eqn:windeq2}) for a wide
 range of these parameters, we found that a transonic solution exists
 for any combination of parameter values.

 In the first part of this section, we discuss the structure of
 viscous transonic decretion disks. In the second part, we discuss
 the topology of the sonic point.
 Since we found that the decretion disk structure is not sensitive to
 the spectral type of the central star, we will present only examples
 in which the central star is a B0 main-sequence star with
 $M=17.8\,M_\solar$, $R=7.41\,R_\solar$, and
 $T_{\rm eff}=2.80 \cdot 10^4\,{\rm K}$
 (Allen 1973), where $T_{\rm eff}$ is the effective temperature.

\subsection{Structure of the Viscous Transonic Outflow}

 Figure~1 shows the structure of the viscous transonic decretion disk
 for $T_{\rm d}={1\over2}T_{\rm eff}$ and $(\eta, \epsilon)=(0,0)$,
 where $T_{\rm d}$ is the disk temperature. In figure~1, solid,
 dashed, and dash-dotted lines denote $V_r/c_{\rm s}$,
 $V_\phi/(GM/R)^{1/2}$, and $\Sigma/\Sigma(R)$, respectively. For
 comparison purpose, transonic solutions for $\alpha=1$ (thick lines),
 $\alpha=0.1$ (lines with intermediate thickness), and $\alpha=0.01$
 (thin lines) are shown.

\begin{figure}[t]
   \hspace*{0.2cm}
   \psfig{file=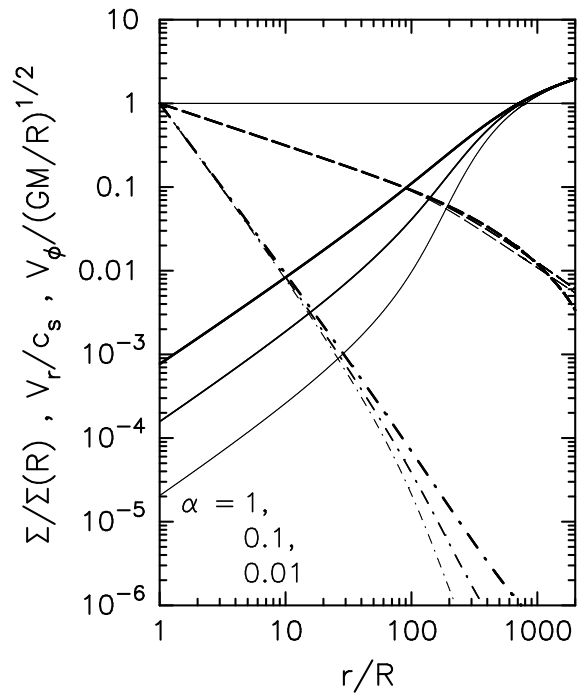,width=8.5cm}
   \vspace*{-1cm}
   \caption[ ]{Structure of
      the viscous transonic decretion disk for $T_{\rm
        d}={1\over2}T_{\rm eff}$ and $(\eta, \epsilon)=(0,0)$. The
      central star is a B0 main-sequence star.  Solid, dashed, and
      dash-dotted lines denote $V_r/c_{\rm s}$, $V_\phi/(GM/R)^{1/2}$,
      and $\Sigma/\Sigma(R)$, respectively.  Thick lines, lines with
      intermediate thickness, and thin lines are for $\alpha=$ 1, 0.1,
      and 0.01, respectively.}
   \label{fig:ts1}
\end{figure}

\begin{figure*}[t]
   \hspace*{0.5cm}
   \psfig{file=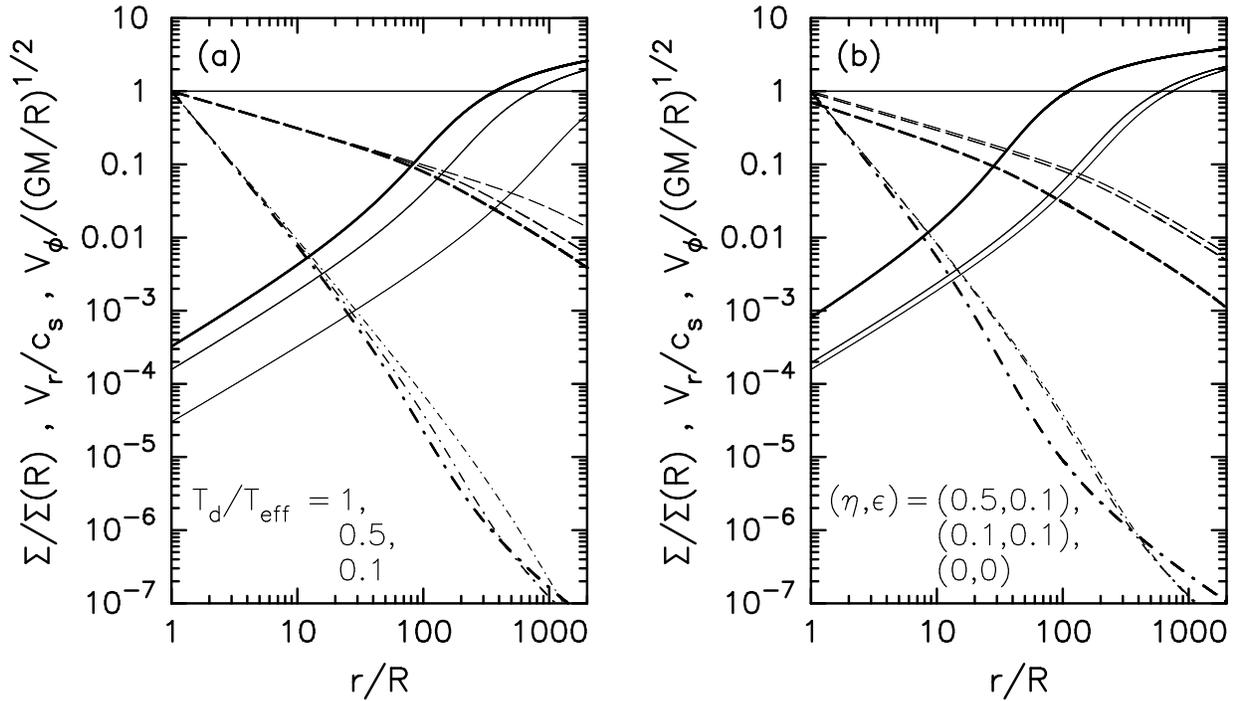,width=17cm,angle=-90.0}
   \vspace*{-2cm}
   \caption[ ]{Parameter dependence of the viscous transonic
     solutions. The value of $\alpha$ is fixed to be 0.1.  The central
     star is a B0 main-sequence star.  (a) The temperature dependence:
     Thick lines, lines with intermediate thickness, and thin lines
     are for $T_{\rm d}/T_{\rm eff}=$ 1, 1/2, and 1/10, respectively.
     No radiative force is included.  (b) The dependence on the
     radiative parameters:  Thick lines and lines with intermediate
     thickness are for $(\eta,\epsilon) = (0.5, 0.1)$ and
     $(\eta,\epsilon) = (0.1, 0.1)$, respectively.  Thin lines are for
     no radiative force.  The disk temperature is fixed to be $T_{\rm
       d}/T_{\rm eff} = 1/2$.  Other format of the figure is the same
     as that of figure~1.}
   \label{fig:ts2-3}
\end{figure*}

 From figure~1, we find that the sonic point is located far from the
 star and the outflow is highly subsonic for $r \ll 10^2\,R$. We also
 find that physical quantities vary approximately as $V_r
 \propto r$, $\Sigma \propto r^{-2}$ (i.e., the local density $\propto
 r^{-7/2}$ in the isothermal disk), and $V_\phi \propto r^{-1/2}$ in
 the inner subsonic part. For smaller $\alpha$, the slopes of $V_r$
 and $\Sigma$ become steeper in the outer subsonic part.  In the outer
 subsonic part and in the supersonic part, $V_\phi$ approximately
 decreases as $r^{-1}$. It is important to note that the transonic
 decretion disk has the transitional feature mentioned in section~1:
 It is near Keplerian in the inner part, while it is angular momentum
 conserving in the outer part. 
 As shown below, these features of viscous transonic decretion disks
 hold irrespective of parameter values.

 As mentioned above, there are five parameters that characterize
 the current problem. In figure~1 we have studied how the viscosity
 parameter affects the transonic solution with the other four
 parameters being fixed.  In figure~2, we show how transonic solutions
 depend on (a) the disk temperature $T_{\rm d}/T_{\rm eff}$ and (b)
 the radiative parameters $\eta$ and $\epsilon$.
 From figure~2, we note that the characteristics of the transonic
 solutions are always the same as those shown in figure~1. We also
 note that the sonic radius is smaller for higher disk temperature
 and/or larger radiative force.

 It is important to note that
 it is basically the pressure force that accelerates the flow
 up to a supersonic speed. The acceleration by the pressure force
 does not work effectively in the region where it is much weaker than
 the effective gravity. This is why the sonic point is located far
 from the star. From equation~(\ref{eqn:windeq1}), the sonic
 radius $r_{\rm s}$ is roughly given by $r_{\rm s} \sim
 {2\over5}GM/c_{\rm s}^2$ when the radiative force is much weaker
 than gravity. That the pressure force plays an important role for the 
 acceleration suggests that the flow structure can be sensitive to the 
 radial distribution of the disk temperature. We have
 assumed, for simplicity, the disk to be isothermal, but the disk
 temperature should be originally determined by solving the energy
 equation. If the disk is not isothermal, it will affect the flow
 structure, in particular in the region of $r>100R$, where the sonic
 point in the isothermal disk lies.

 We would like to make a brief comment here on the radial velocity of
 viscous decretion disks. One may estimate from
 equation~(\ref{eqn:angmom}) that the radial velocity $V_r$ is of the
 order of $\alpha (H/r)^2 V_\phi$, which is reduced to
 $0.6 \alpha (T_{\rm d}/T_{\rm eff}) (r/R)^{1/2}\,{\rm km\,s}^{-1}$
 for an isothermal, near Keplerian disk around a B0-type main-sequence
 star. As shown in figures~1 and 2, however, this estimate is wrong in
 both absolute magnitude and radial dependence.
 Equation~(\ref{eqn:windeq2}) indicates that $V_r$ roughly
 linearly increase with $r$, as discussed at the end of section~2,
 and has to be much smaller than $\alpha (H/r)^2 V_\phi$ near the
 star.

\subsection{Topology of the Sonic Point}

 It is well known that the stability of the transonic accretion
 with Shakura-Sunyaev viscosity prescriptions
 is related to the topology of the sonic point
 (e.g., Abramowicz, Kato 1989, and references therein).
 For $\alpha$ smaller than a critical value,
 the sonic point is of saddle type and the transonic accretion
 is stable. On the other hand, for $\alpha$ greater than the critical value,
 the sonic point is nodal and
 the transonic accretion becomes unstable.
 The instability arises as a result of work done by viscous force
 [see Kato et al.\ (1988) for detailed analysis].

\begin{figure*}[t]
   \hspace*{0.5cm}
   \psfig{file=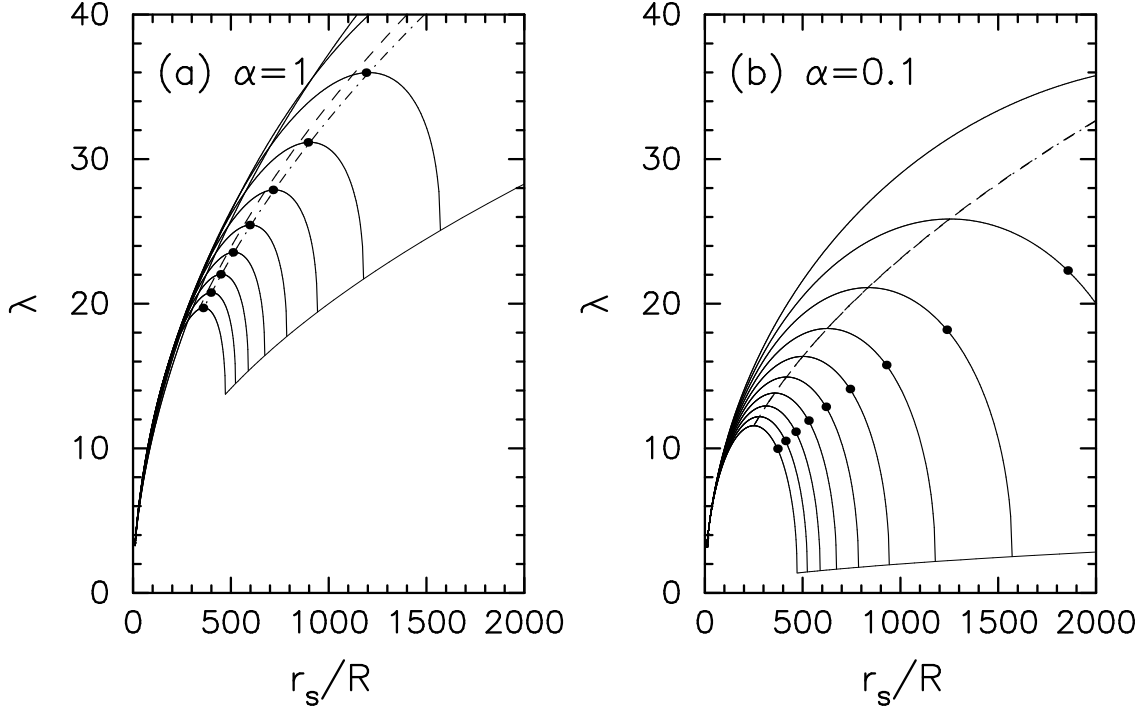,width=16cm,angle=-90.0}
   \vspace*{-1.5cm}
   \caption[ ]{Diagram showing the location of the sonic points
     in the $(r_{\rm s}, \lambda_{\rm s})$-plane for (a) $\alpha=1$
     and (b) $\alpha=0.1$. No radiative force is included and
     $\lambda_{\rm s} = \ell_{\rm s}+\alpha c_{\rm s} r_{\rm s}$ is
     normalized by $(GMR)^{1/2}$. The central star is a B0-type
     main-sequence star.  Each contour denotes the relation between
     $r_{\rm s}$ and $\lambda_{\rm s}$ for a constant sound speed.
     From the bottommost curve upwards, $T_{\rm d}/T_{\rm eff}$ $=$ 1,
     0.9, 0.8, $\ldots$, 0.3, 0.2, and 0.1.  The filled circle on each
     contour indicates the position of the sonic point.  In each
     panel, the dashed line separates the spiral-type sonic point
     region (upper side of the line) from the nodal-type sonic point
     region (lower side of the line), and the dash-dotted line
     separates the nodal-type region (upper side) from the saddle-type
     region (lower side).  In figure~3b the nodal-type region is so
     narrow that it is indistinguishable on the figure.  The thin
     solid line in each panel indicates a boundary in the $(r_{\rm s},
     \lambda_{\rm s})$-plane, below which the circular orbit at the
     sonic radius is unstable.}
   \label{fig:topol}
\end{figure*}

 In the current model, both left-hand and right-hand sides
 of equation~(\ref{eqn:windeq1}) must vanish at the sonic point, i.e.,
 \begin{equation}
    V_r(r_{\rm s}) = c_{\rm s},
    \label{eqn:sonic1}
 \end{equation}
 \begin{equation}
    \ell_{\rm s}^2 = GMr_{\rm s}
         \left[1-\eta\left({r_{\rm s} \over R}\right)^\epsilon\right]
         -{5\over2}c_{\rm s}^2 r_{\rm s}^2.
    \label{eqn:sonic2}
 \end{equation}
 To examine the topology of the sonic point, we evaluate the
 derivative ${\rm d}V_r/{\rm d}r$ at $r=r_{\rm s}$. After some
 manipulations, we have
 \begin{equation}
    \left. {{{\rm d}V_r} \over {{\rm d}r}} \right|_{r_{\rm s}} =
         {{\alpha \ell_{\rm s}} \over {2r_{\rm s}^2}}
         \pm \left( {{\alpha^2 \ell_{\rm s}}^2 \over {4r_{\rm s}^4}}
         - X \right)^{1/2},
    \label{eqn:reg1}
 \end{equation}
 where
 \begin{eqnarray}
    X &=& -{{2GM} \over {r_{\rm s}^3}}
         \left[1-\eta(2-\epsilon)
         \left({r_{\rm s} \over R}\right)^\epsilon\right]
         \nonumber \\
      && +{3\over2}{\ell_{\rm s}^2 \over r_{\rm s}^4}
         +{{\alpha c_{\rm s} \ell_{\rm s}} \over r_{\rm s}^3}
         +{5\over4}{c_{\rm s}^2 \over r_{\rm s}^2}.
    \label{eqn:reg2}
 \end{eqnarray}
 Note that equations~(\ref{eqn:reg1}) and (\ref{eqn:reg2}) are
 quite similar to equations~(2.15) and (2.16) of Abramowicz and Kato
 (1989) for the transonic accretion.
 According to the standard classification of critical points
 (e.g., Ferrari et al.\ 1985),
 the sonic point is of saddle type if $X<0$,
 nodal if $0<X<\alpha^2\ell_{\rm s}^2/4r_{\rm s}^4$,
 and spiral if $X>\alpha^2\ell_{\rm s}^2/4r_{\rm s}^4$.

 In figure~3, as typical examples, we present the location of the
 sonic points for (a) $\alpha=1$ and (b) $\alpha=0.1$. In the figure,
 $\lambda_{\rm s} \equiv \ell_{\rm s}+\alpha c_{\rm s} r_{\rm s}$ is
 normalized by $(GMR)^{1/2}$. Solid contours denote the relation
 between $r_{\rm s}$ and $\lambda_{\rm s}$, given by
 equation~(\ref{eqn:sonic2}), for $T_{\rm d}/T_{\rm eff} =$ 1, 0.9,
 0.8, $\ldots$, 0.3, 0.2, and 0.1 from the bottom to the top.  The
 filled circle on each contour indicates the position of the sonic
 point.  In each panel, the dashed line separates the spiral-type
 sonic point region (upper side of the line) from the nodal-type sonic
 point region (lower side of the line), and the dash-dotted line
 separates the nodal-type region (upper side) from the saddle-type
 region (lower side).  In figure~3b the nodal-type region is so narrow
 that it is indistinguishable on the figure.  The thin solid line in
 each panel indicates a boundary in the $(r_{\rm s}, \lambda_{\rm
 s})$-plane, below which the circular orbit at the sonic radius is
 unstable.

 Figure~3 shows that the sonic point for $\alpha=1$
 is located in the nodal-type region,
 whereas that for $\alpha=0.1$ is in the saddle-type region.
 Examining the transonic solutions for a wide range of parameters,
 we found that the topology of the sonic point depends only on $\alpha$
 and there is a critical value, $\alpha_*$, such that
 the sonic point is nodal for $\alpha>\alpha_*$, while
 it is of saddle type for $\alpha<\alpha_*$.
 Note the similarity between the transonic decretion and accretion.
 In our model, $\alpha_*$ lies between 0.9 and 0.95.
 This means that the steady transonic decretion in disks of Be stars
 exists for $\alpha \ltsim 0.9$, whereas the transonic decretion
 cannot be steady for $\alpha \gtsim 0.95$.

\section{Summary and Discussion}\label{sec:summary}

 We have examined the characteristics of the outflow in
 viscous decretion disks around Be stars.
 For simplicity, we have assumed the disk to be isothermal
 and adopted the Shakura-Sunyaev $\alpha$-prescription for the viscous
 stress. As the radiative force,
 we have adopted the parametric form used by Chen and Marlborough (1994).

 Solving the resulting wind equations
 for a wide range of parameter values,
 we have found that a transonic solution exists for any combination
 of parameters.
 When the radiative force is much smaller than the gravity,
 it is basically the pressure force
 that accelerates the flow up to a supersonic speed.
 The sonic radius $r_{\rm s}$ is, then, roughly given by
 $r_{\rm s} \sim {2\over5}GM/c_{\rm s}^2$.
 The sonic radius is smaller for higher disk temperature.
 A larger radiative force also gives a smaller sonic radius.
 The outflow is highly subsonic
 in the inner part of the disk.
 Approximately, the radial velocity increases as $r$ and
 the surface density decreases as $r^{-2}$ in the inner subsonic part.
 For smaller $\alpha$,
 slopes of these quantities become steeper in the outer subsonic part.
 Interestingly, the disk is near Keplerian ($V_\phi \sim r^{-1/2}$)
 in the inner subsonic part, while
 it is angular momentum conserving ($V_\phi \sim r^{-1}$)
 in the outer subsonic part and in the supersonic part.

 We have also found that
 the topology of the sonic point is nodal for $\alpha$ greater than
 a critical value, $\alpha_*$,
 whereas it is of saddle type for $\alpha < \alpha_*$.
 In our model, $\alpha_*$
 lies between 0.9 and 0.95
 irrespective of values of other parameters.
 From the similarity between the physics of accretion and decretion and
 the study of the transonic accretion disks, we expect that the transonic
 decretion disks with $\alpha < \alpha_*$ are stable, whereas
 those with $\alpha > \alpha_*$ are unstable.

 We have seen that the viscous transonic decretion disks have many
 characteristics in agreement or consistent with the observed
 features. Thus, the current model is satisfactory as a first step to
 obtain a sophisticated model of Be disks. We have to admit, however,
 that there are several issues which remain unsolved by the current
 version of the model. They are the observed high mass-loss rate
 through the disk or short disk-formation timescale, a variety of the
 observed density distribution, and
 the mass supply mechanism from the star. We would like to discuss
 these issues in the rest of this paper.

 First, we consider the mass loss rate $|\dot{M}|$ through the disk.
 In principle, the mass loss rate is determined by how much torque is
 exerted by the star at the inner disk boundary.
 Unfortunately, however, we have no satisfactory model for the
 star-disk interaction or observational estimate of $|\dot{M}|$ in
 quasi-steady disks ($|\dot{M}|$ in the disk formation stage will be
 discussed later). Therefore, in the below we give an indirect
 estimate of $|\dot{M}|$ the present model for the steady, viscous
 transonic solution allows.

 In isothermal decretion, as in isothermal accretion,
 $\dot{M}$ is not a parameter which characterizes the problem.
 It does not appear in the wind equations~(\ref{eqn:windeq1}) and
 (\ref{eqn:windeq2}),
 so the absolute value of the surface density $\Sigma$ is not
 determined from the model.
 Using the observed base density (i.e., the equatorial density
 at the inner disk radius $r=R$) to normalize $\Sigma$, however,
 we can estimate $\dot{M}$ by
 \begin{eqnarray}
    |\dot{M}| & \sim & 2 \times 10^{-11} \cdot
    {{\rho_{00}(R)} \over {10^{-11} {\rm g}\,{\rm cm}^{-3}}} \cdot
    {{V_r(R)} \over {10^{-3} c_{\rm s}}}
    \nonumber \\
    && \times
    \left( {T_{\rm d} \over T_{\rm eff}} \right)^{1/2}
    M_\solar\,{\rm yr}^{-1}.
    \label{eq:Mdot}
 \end{eqnarray}
 The base density, $\rho_{00}(R)$, for most stars analyzed from
 Fe~II lines (Hanuschik 1986) and the IR excess
 (Waters et al.\ 1987; Dougherty et al.\ 1994)
 ranges from $10^{-11}$ to $10^{-12}$ g\,cm$^{-3}$.
 Figure~1 shows that the radial velocity at the inner disk
 radius of the transonic decretion disks, $V_r(R)$, is roughly of the
 order of $10^{-3} \alpha c_{\rm s}$, as far as the radiative force is
 much smaller than the gravity.
 As a result, we obtain $|\dot{M}|$
 of the order of $10^{-11} - 10^{-12} \alpha M_\solar\,{\rm yr}^{-1}$
 for steady, viscous transonic disks.

 It should be noted that this value does not increase very much even
 if we use $V_r(R)$ derived from the observed upper limit of the disk
 outflow velocity, i.e., a few km\,s$^{-1}$ for $r \ltsim
 10R$. Since $V_r$ approximately linearly increases with $r$ in
 viscous decretion disks, this upper limit of the disk outflow,
 $V_r(10R) <$ a few km\,s$^{-1}$ $\sim 0.1 c_{\rm s}$, suggests that
 $V_r(R) \ltsim 10^{-2} c_{\rm s}$ and $r_{\rm s} \gtsim 10^2
 R$. Consequently, using the upper limit of the observed radial flow,
 we have $|\dot{M}| \sim 10^{-10} - 10^{-11} M_\solar\,{\rm yr}^{-1}$
 from equation~(\ref{eq:Mdot}). Our results, together with the
 observed range of the base density of Be disks, suggest that, in
 steady viscous decretion, the mass loss rate through the disk is at
 most comparable with that in the polar wind region.

 It is interesting to note that another constraint on $\dot{M}$ is
 obtained by estimating the spin-down timescale of the Be
 star. Porter~(1998) derived the constraint on a quantity equivalent
 to $\dot{M}$, for which Be stars do not show significant spin-down
 during their main-sequence lifetimes. Porter adopted that each Be
 star loses its angular momentum at the rate of $|\dot{M}| \ell_{\rm
 K}(R)$, where $\ell_{\rm K}(R)$ is the Keplerian angular momentum at
 $r=R$ (here, for the sake of convenience, we have adopted the
 notation different from Porter's). In viscous disks, however, this
 rate is largely an underestimate, because it is likely that the
 Keplerian region extends at least up to $100R$ where the specific
 angular momentum is one order of magnitude larger than $\ell_{\rm
 K}(R)$. Thus, we have to use $\sim 10|\dot{M}| \ell_{\rm K}(R)$
 instead of $|\dot{M}| \ell_{\rm K}(R)$ to evaluate the
 angular-momentum loss rate from the star. Then, the constraint for
 the star not to show significant spin-down during their lifetimes
 becomes more stringent than that derived by Porter (1998).

 Since it is beyond the scope of the paper to present a detailed
 calculation, we roughly estimate the spin-down timescale below. For
 this purpose, we adopt the same stellar model as that adopted by
 Porter, i.e., a rigidly rotating polytrope with the index of 3/2. The
 stellar angular momentum is, then, given by $\sim 0.2 f M \ell_{\rm
 K}(R)$, where $f$ is the ratio of the stellar rotation velocity to
 the break-up velocity, which ranges from 0.4 for early Be stars to
 0.8 for late Be stars. Then, the spin-down timescale is given by
 $\sim 0.2 f M \ell_{\rm K}(R)/10|\dot{M}| \ell_{\rm K}(R)$ $\sim 0.02
 f M/|\dot{M}|$ $\sim 10^9 (|\dot{M}|/10^{-10}M_\solar\,{\rm
 yr}^{-1})\,{\rm yr}$. Since the spin-down timescale has to be much
 longer than the main-sequence lifetime, which is $\sim 10^7\,{\rm
 yr}$ for early Be stars and $\sim 10^8\,{\rm yr}$ for late Be stars,
 we have the constraint on $\dot{M}$ for Be stars: $|\dot{M}| \ll
 10^{-8}M_\solar\,{\rm yr}^{-1}$ for early Be stars and $|\dot{M}| \ll
 10^{-9}M_\solar\,{\rm yr}^{-1}$ for late Be stars. Since the observed
 disk structure of Be stars is insensitive to the spectral type, the
 latter constraint ($|\dot{M}| \ll 10^{-9}M_\solar\,{\rm yr}^{-1}$)
 should be taken.

 Observationally, the rate of equatorial mass loss has been derived
 for several stars only in the disk formation stage.
 Since 1977, $\mu$~Cen has exhibited only flickering emission caused by
 episodic mass loss events. Each outburst lasts only for 2--5\,d and
 makes a transient, small disk which decays in 20--80\,d. Hanuschik et
 al.\ (1993) estimated the mass loss rate for $\mu$~Cen during each
 outburst to be $4 \times 10^{-9} M_\solar\,{\rm yr}^{-1}$. Recently,
 Telting et al.\ (1998) studied the long-term behavior of the Be disk
 in X Per (4U~0352$+$30), a Be/X-ray binary system, and derived the
 mass loss rate of $4 \times 10^{-9} M_\solar\,{\rm yr}^{-1}$ during a
 disk build-up phase which lasted about 1\,y in 1994--1995.

 These values show that Be stars can eject mass at a rate much higher
 than the upper limit predicted by the viscous decretion model for
 steady disks. This suggests that the disk formation is a violent and,
 consequently, non-steady process to which the the steady disk model
 cannot be applied. To study such a process, we need numerical
 simulations such as those by Kroll and Hanuschik (1997).

 The short timescale of disk formation around these stars also
 suggests that the disk-formation process is essentially
 non-steady. For steady, viscous transonic disks shown in figure~1,
 the drift time defined by $\int_R^r V_r^{-1} dr$ is $\sim 30
 V_r(R)/10^{-3}c_{\rm s}\,{\rm yr}$ at $r=10R$ and $\sim 80
 V_r(R)/10^{-3}c_{\rm s}\,{\rm yr}$ at the sonic radius. This means,
 even if $V_r(R)$ is as fast as $10^{-2}c_{\rm s}$, the formation of
 an $r \sim 10R$ disk takes about 3\,yr and that of a transonic disk
 takes about 8\,yr. This long drift time suggests that only disks
 which are persistent for more than 10\,yr will have the structure
 similar to those studied in this paper.

 Next, we discuss the density distribution in Be disks.
 In isothermal decretion disks,
 the radial exponent of density, $n$, defined by $n=-d \ln \rho_{00}/d
 \ln r = -d \ln (\Sigma/H)/d \ln r$, is a little bit bigger than
 3.5. More generally, as pointed out by Porter (1999),
 $2n+3m \gtsim 7$ must hold in order for the decretion to occur, where
 $m$ is the radial exponent of the disk temperature. Since Be disks are
 roughly isothermal except for the very vicinity of the star (Millar,
 Marlborough 1998, 1999), we would expect that the observed density
 gradient index $n$ is distributed around $n \sim 3.5$ with a small
 dispersion.

 On the other hand, the observed values of $n$ are distributed around
 $n\sim3-3.5$, which is in agreement with the current model, with a
 dispersion much wider than that predicted. Adopting a simplified disk
 model, in which the disk has a constant opening angle and the density
 depends only on radius, and applying the curve of growth method by
 Waters (1986) to the observed far-IR excess for 59 Be stars, Waters et
 al.\ (1987) found $n = 2-4$. Adopting the same disk model as that of
 Waters et al.\ (1987), Dougherty et al.\ (1994) obtained $n=2-5$ from
 the near-IR excess colors for 144 Be stars. Moreover, applying the
 curve of growth method with a more realistic disk model to a Be star,
 Porter (1999) obtained a value of $n$ quite similar to that obtained
 by Waters et al.\ (1987).

 Consequently, the viscous decretion disk model is roughly in
 agreements with the observed density distribution, but it has to
 explain why the observed density distribution ranges widely. In
 particular, the model has to explain why there are Be stars with $n
 \ltsim 3$, for which the decretion does not seem to occur. Porter
 (1999) pointed out that the viscosity dependent on the radius and/or
 the stellar radiation field may help make a decretion disk with a
 shallow density gradient. More detailed study of these possibilities
 is desired.
 
 Finally, we come to the most important and long-standing issue to be
 solved, that is, the mechanism which enables a star to supply mass
 into the surrounding disk. Be stars rotate significantly slowly
 compared with the break-up velocities. Therefore, in order to supply
 mass, stars have to accelerate matter up to break-up
 velocities. Moreover, since most Be stars keep their disks for long
 periods, either the mass supply must be continuous or the
 interval of mass supply events must be much shorter than the viscous
 timescale. Otherwise, the disk would decay before the next mass supply
 event occurs, like the disk of $\mu$~Cen.

 Although there is no widely-accepted mechanism for such mass supply,
 non-radial pulsations of stars may work. Osaki (1986) suggested that
 the dissipation of the non-radial pulsations gives matter the
 necessary angular momentum. Recently, Rivinius et al.\ (1997) found
 that in $\mu$~Cen the time interval of mass supply events is equal to
 a beat period of several strongest non-radial pulsation (NRP) modes,
 indicating that the mass supply from the star results from the
 resonance between strongest NRP modes. If this mechanism works in
 other Be stars, the frequencies of the strongest NRP modes in most Be
 stars will be distributed so that the beat period becomes much shorter
 than the viscous timescale. It is highly desirable to study this
 possibility further.

 \vspace{1pc}\par

 The author is grateful for H.F.\ Henrichs, I.\ Negueruela, and J.M.\
 Porter for stimulating discussions. He thanks the anonymous referee
 for useful comments.
 He also thanks the Astronomical Institute \lq Anton Pannekoek',
 University of Amsterdam, for the warm hospitality,
 where during his stay most of this work was performed.

\section*{References}

{\small
\re
Abott, D.C.\ 1982, ApJ 259, 282
\re
Abramowicz, M.A., Kato, S.\ 1989, ApJ 336, 304
\re
Allen, C.W.\ 1973, in Astrophysical Quantities, 3rd ed
     (The Athlone Press) London,  pp206, 209
\re
Castor, J.I., Abott, D.C., Klein, R.I.\ 1975, ApJ 195, 157
\re
Chen, H., Marlborough, J.M.\ 1994, ApJ 427, 1005
\re
Dougherty, S.M., Waters, L.B.F.M., Burki, G., Cot\'{e}, J., Cramer, N.,
      Van Kerkwijk, M.H., Taylor, A.R.\ 1994, A\&A 290, 609
\re
Ferrari, A., Trussoni, E., Rosner, R., Tsinganos, K.,
      1985, ApJ 294, 397
\re
Hanuschik, R.W.\ 1987, A\&A 173, 299
\re
Hanuschik, R.W.\ 1994, in Pulsation, Rotation and Mass Loss
      of Early-Type Stars, IAU Symp 162,
      ed L.A.\ Balona, H.\ Henrichs, J.M.\ Le Contel
      (Kluwer Academic Publishers, Dordrecht) p265
\re
Hanuschik, R.W.\ 2000, in The Be Phenomenon in
      Early-Type Stars, IAU Colloq 175,
      ed M.\ Smith, H.F.\ Henrichs, J.\ Fabregat
      (ASP, San Francisco) in press
\re
Hanuschik, R.W., Dachs, J., Baudzus, M., Thimm, G.\ 1993, A\&A 274, 356
\re
Hanuschik, R.W., Hummel, W., Dietle, O., Sutorios, E., 1995, A\&A 300, 163
\re
Kato, S., Honma, F., Matsumoto, R.\ 1988,
      MNRAS 231, 37
\re
Kroll, P., Hanuschik, R.W.\ 1997, in Accretion Phenomena and
      Related Outflows, IAU Colloq 163,
      ed D.T.\ Wickramasinghe, G.V.\ Bicknell, L.\ Ferrario
      (ASP, San Francisco) p494
\re
Lee, U., Saio, H., Osaki, Y.\ 1991, MNRAS 250, 432
\re
Matsumoto, R., Kato, S., Fukue, J., Okazaki, A.T.,
      1984, PASJ 36, 71
\re
Millar, C.E., Marlborough, J.M.\ 1998, ApJ 494, 715
\re
Millar, C.E., Marlborough, J.M.\ 1999, ApJ 516, 276
\re
Negueruela, I., Okazaki, A.T.\ 2000, A\&A, accepted
\re
Okazaki, A.T.\ 1991, PASJ 43, 75
\re
Osaki, Y.\ 1986, PASP 98, 30
\re
Papaloizou J.C., Savonije G.J., Henrichs H.F.\ 1992, A\&A 265, L45
\re
Porter, J.M.\ 1998, A\&A 333, L83
\re
Porter, J.M.\ 1999, A\&A 348, 512
\re
Quirrenbach, K.S., Bjorkmann, K.S., Bjorkman, J.E., Hummel, C.A.,
       Buscher, D.F., Armstrong, J.T., Mozukewich, D., Elias, N.M.II.,
       Babler, B.L.\ 1997, ApJ 479, 477
\re
Rivinius, Th., Baade, D., \v{S}tefl, S., Stahl, O., Wolf, B., Kaufer, A.\ 1997,
       in A Half Century of Stellar Pulsation Interpretations:
       A Tribute to Arthur N.\ Cox, ed P.A.\ Bradley, J.A.\ Guzik
       (ASP, San Francisco) p343
\re
Snow, T.P.\ 1981, ApJ 251, 139
\re
Telting, J.H., Waters, L.B.F.M., Roche, P., Boogert, A.C.A., Clark, J.S.,
       de Martino, D., Persi, P.\ 1998, A\&A 296, 785
\re
Waters, L.B.F.M.\ 1986, A\&A 162, 121
\re
Waters, L.B.F.M., Cot\'{e}, J., Lamers, H.J.G.L.M.\ 1987, A\&A 185,
      206
\re
Waters, L.B.F.M., Marlborough, J.M.\ 1994,
      in Pulsation, Rotation and Mass Loss
      of Early-Type Stars, IAU Symp 162,
      ed L.A.\ Balona, H.\ Henrichs, J.M.\ Le Contel
      (Kluwer Academic Publishers, Dordrecht) p399
\re
}


\end{document}